\documentclass[acmlarge, nonacm]{acmart}
\AtBeginDocument{%
  }

\setcopyright{none}
\renewcommand\footnotetextcopyrightpermission[1]{}

\begin{document}

\title{Audio-Augmented Sensory Technology for Everyday 
Mindfulness: An In-the-Wild Diary Study}

\author{Yujie Tao}
\email{yjtao@stanford.edu}
\orcid{0000-0002-3492-6747}
\affiliation{%
  \institution{Stanford University}
  \city{Stanford}
  \state{California}
  \country{USA}
}

\author{Jingjin Li}
\email{jingjin@aimpower.org}
\orcid{0000-0003-0193-7228}
\affiliation{%
  \institution{AImpower.org}
  \city{Mountain View}
  \state{California}
  \country{USA}
}

\author{Bruce Ruan}
\email{bjruan@stanford.edu}
\orcid{0009-0004-5552-671X}
\affiliation{%
  \institution{Stanford University}
  \city{Stanford}
  \state{California}
  \country{USA}
}

\author{Hui Qi Tong}
\email{htong@stanford.edu}
\orcid{0000-0002-9121-2580}
\affiliation{%
  \institution{Stanford University}
  \city{Stanford}
  \state{California}
  \country{USA}
}

\author {Jeremy N. Bailenson}
\email{bailenso@stanford.edu}
\orcid{0000-0003-2813-3297}
\affiliation{%
  \institution{Stanford University}
  \city{Stanford}
  \state{California}
  \country{USA}
}

\author{Sean Follmer}
\email{sfollmer@stanford.edu}
\orcid{0000-0001-5592-5949}
\affiliation{%
  \institution{Stanford University}
  \city{Stanford}
  \state{California}
  \country{USA}
}



\begin{abstract}

Bringing present-moment awareness to routine activities, from cooking to brushing teeth, is a mindfulness practice commonly incorporated into evidence-based mental health intervention programs. However, practicing everyday mindfulness can be challenging, as it requires people to pay deliberate attention to actions they typically perform on autopilot. Audio-augmented sensory technology offers a promising way to support mindfulness by amplifying sounds produced by ongoing manual interactions, from the scrape of a wooden spoon against the bottom of a pot to the soft scratch of a pen moving across paper. While prior work has shown promising laboratory findings, the use of audio augmentation in everyday settings remains underexplored. In this work, we conducted a week-long diary study with 12 experienced mindfulness practitioners to examine how audio augmentation of manual interactions affects everyday mindfulness practice. Participants contributed nearly 90 hours of device use, 84 diary entries, and 12 post-study interviews. Participants reported that the system supported their mindfulness practice over multiple days and the usage depended on task demands and auditory characteristics. We identified roles of the system in scaffolding mindfulness practice, such as easing the transition into a mindful state and providing experiential targets for long-term practice.
\end{abstract}

\begin{teaserfigure}
  \includegraphics[width=\textwidth]{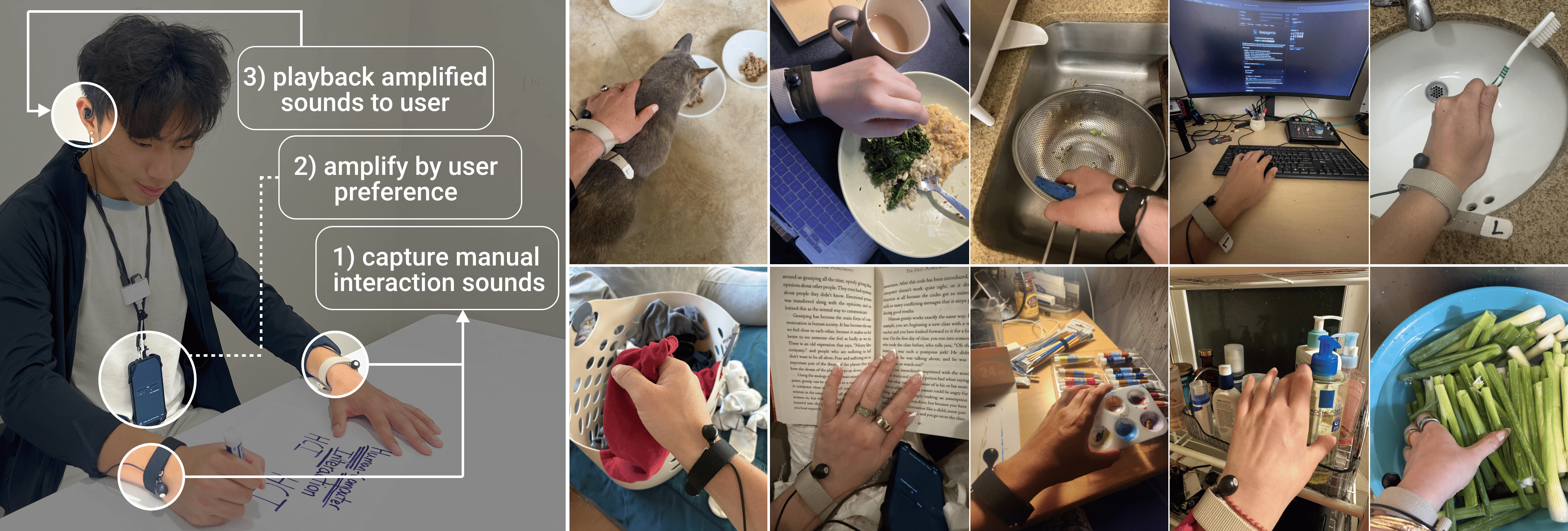}
  \caption{Left: A posed demonstration of the audio augmentation device studied in the week-long diary study. It captures, amplifies and plays back sounds produced by users' hand interactions to enhance present-moment awareness. Right: Photos provided by diary study participants documenting device use across a variety of daily activities. Top row (left to right): petting a cat, eating, washing dishes, working on a desktop, and brushing teeth. Bottom row (left to right): folding laundry, reading a book, painting, skincare, and washing vegetables.}
  \Description{}
  \label{fig:teaser}
\end{teaserfigure}




\maketitle

\section{Introduction}
Mindfulness, the practice of attending to present-moment experience with curiosity and openness, is central to a number of effective mental health interventions~\cite{creswell2017mindfulness}. Established programs such as Mindfulness-Based Stress Reduction~\cite{kabat2003mindfulness} and Mindfulness-Based Cognitive Therapy~\cite{kuyken2010does} have shown clinical benefits for conditions such as anxiety disorder~\cite{hoge2013original}, depression ~\cite{kuyken2015effectiveness}, ADHD~\cite{cairncross2020effectiveness}, substance use disorders~\cite{chiesa2014mindfulness}, and eating disorders~\cite{courbasson2010mindfulness}. Everyday mindfulness, or informal mindfulness, offers a flexible way of practicing by encouraging open awareness of present-moment experiences during daily activities~\cite{hanh2016miracle, creswell2017mindfulness, birtwell2019exploration}. This could be noticing the sound of running water while washing hands or listening to the gentle mixing of ingredients while cooking, combining with an attitude of curiosity, openness, and acceptance. Prior research has shown that everyday mindfulness practice frequency is a strong predictor of psychological flexibility and well-being~\cite{birtwell2019exploration, kakoschke2021importance}.

Despite these benefits, practicing mindfulness during everyday activities can be challenging, as it requires users to bring intentional awareness to familiar routines often performed on autopilot~\cite{li2023co, xie2024understanding}. A growing body of literature calls for technological tools to support everyday mindfulness~\cite{li2024beyond, choi2022design, tao2025audio}, complementing existing systems designed to support formal mindfulness practice, such as meditation. Among these explorations, audio-augmented sensory technology shows great promise. Specifically, Tao et al. proposed to support mindfulness in everyday manual interactions by enabling users to hear sounds produced by their hands amplified in real time~\cite {tao2025audio}. The resulting system aims to ground users' awareness in familiar yet often overlooked sensory experiences, such as opening and closing of a pair of scissors and contact sounds of fingers sliding through different surface textures. An in-lab study with 60 participants showed that audio augmentation of manual interactions enhances state mindfulness and its two core components, attention and curiosity, during object exploration.

While prior work demonstrated the promise of audio augmentation for supporting mindfulness, laboratory studies provided limited insight into how this technique would be integrated in everyday mindfulness practice out-of-lab. Questions therefore remain about the factors that influence its usage in existing practice, the role it plays, and the type of practitioners who could benefit from the system. In this work, we investigate the situated use of audio augmentation of manual interactions in everyday mindfulness practice. Specifically, we developed a portable audio augmentation system compatible with commercial smartphones and conducted a week-long diary study with 12 experienced mindfulness practitioners. Participants were asked to use the device for at least 30 minutes each day during their everyday mindfulness practices and documented their experiences through daily diary questionnaires.

Collectively, 12 participants used the system for 89.84 hours, contributing to 84 diary entries and 298 photos documenting their experiences over a week-long period. Through thematic analysis, we found that participants reported a positive impact of the system on their mindfulness practices across multiple days. This was reflected in participants’ reports of increased awareness of sensory experiences that they would otherwise overlook (e.g., opening and closing pen caps, pouring water from a tap), which in turn influenced their everyday interactions (e.g., slowing down their interactions with everyday objects and using less digital media during mundane activities). 

Moreover, we identified task demand (e.g., when an ongoing task requires substantial focus to complete) and auditory characteristics (e.g., when sounds are already sufficiently prominent) as key factors influencing usage of the device in everyday mindfulness. The system was perceived as suitable for both novice and experienced mindfulness practitioners, although it may serve different roles for each group: novices could benefit from the system by using it to ease the transition into a mindful state or as an experiential target; for experienced practitioners, it could support practice of mindfulness during activities in which maintaining such a state would otherwise be challenging. Despite the device's role in facilitating mindfulness practice, participants emphasized that it should not replace deliberate practice itself. The study findings inform considerations when adopting audio-augmented sensory technology in everyday mindfulness practice and the potential roles it can play in facilitating mindfulness training.

\section{Related Work}
\subsection{Mindfulness and Technology to Support Mindfulness}
Mindfulness is the state of maintaining attention to the present moment with curiosity and openness~\cite{creswell2017mindfulness, bishop2004mindfulness}. Training individuals to cultivate and sustain a mindful state is a central components of several evidence-based mental health interventions~\cite{spijkerman2016effectiveness, creswell2017mindfulness}. Programs such as Mindfulness-Based Stress Reduction (MBSR)~\cite{kabat2003mindfulness} and Mindfulness-Based Cognitive Therapy (MBCT)~\cite{kuyken2010does} have demonstrated robust clinical benefits across a range of conditions, including Generalized Anxiety Disorder (GAD)~\cite{hoge2013original}, recurrent depression ~\cite{kuyken2015effectiveness}, Deficit/Hyperactivity Disorder (ADHD)~\cite{cairncross2020effectiveness}, substance use disorders~\cite{chiesa2014mindfulness}, and eating disorders~\cite{courbasson2010mindfulness}. 

Mobile and wearable technologies designed to support mindfulness have largely focused on formal practices—structured exercises in which individuals intentionally dedicate time and space to practice mindfulness~\cite{birtwell2019exploration, li2024beyond}. These practices commonly include meditation~\cite{moore2009meditation}, yoga~\cite{gaiswinkler2016relationship}, and Tai Chi~\cite{chen2021effects}. Mobile applications such as \textit{Headspace}~\cite{headspace} and \textit{Calm}~\cite{calm} broaden access to expert-guided mindfulness practices through pre-recorded content~\cite{creswell2025meditation}. The rise of large language models enables personalization of meditation instruction to prompt user engagement~\cite{wu2026mindfulagents, kumar2025large, bennardo2026stream}. Live-streaming platforms allow users to practice mindfulness together despite being physically apart~\cite{li2024meditating,islam2025verbal}, and virtual reality further enhances co-presence of remote users in the same 3D spaces~\cite{liu2025meditating, feinberg2022zenvr}. Mixed reality glasses enable visualization of inner states layered on top of physical objects~\cite{roo2017inner} and, beyond visual feedback, novel multimodal cues such as auditory~\cite{tan2025runme}, tactile~\cite{ezer2024somaesthetic}, and olfactory feedback~\cite{xia2026design} have been designed to guide attention during meditation. RunMe~\cite{tan2025runme}, moreover, proposed a closed-loop system that adaptively delivers auditory guidance based on users’ ongoing states during running meditation.

\subsection{Interactive Systems to Support Everyday Mindfulness}
Beyond formal practices, mindfulness can also take the form of informal practices, or everyday mindfulness, where present-moment awareness is woven into everyday activities~\cite{hanh2016miracle,thompson2007everyday, terzimehic2019review, shankland2021improving}. This can involve cultivating mindful attention during routine activities as simple as washing dishes~\cite{hanley2015washing}, folding clothes~\cite{hanh2016miracle}, or eating~\cite{warren2017structured}. Unlike formal practices, everyday mindfulness is less structured and does not require users to dedicate separate time and space for practice. Prior work has shown that the frequency of everyday mindfulness practice is a strong predictor of psychological flexibility and well-being~\cite{birtwell2019exploration, kakoschke2021importance}.

Although everyday mindfulness allows people to integrate practice into activities they already engage in~\cite{hanley2015washing, hanh2016miracle}, sustaining mindful awareness can be challenging because these activities are habitually performed on autopilot~\cite{li2023co, xie2024understanding}. The dynamic nature of everyday contexts adds to this challenge, as competing stimuli can divert practitioners’ attention and make it difficult to maintain consistent awareness~\cite{osin2022mindfulness, li2024beyond}.

Recent years have seen growing attention to this challenge, where researchers call for technologies to support mindfulness beyond formal practice. For example, Li et al.~\cite{li2023co} conducted a design workshop with 30 mindfulness practitioners to reflect on personal experiences of everyday mindfulness and to explore designs that support daily practice. To support everyday mindfulness, much work focuses on how to deliver guided instruction unobtrusively. Tan et al.~\cite{tan2023mindful}, for instance, explored different visualization patterns for integrating mindfulness exercises with smart glasses to enable “on-the-go” practice. Choi et al.~\cite{choi2022design} integrated subtle tactile cues with everyday objects such as a seatbelt to guide breathing in everyday activities. Gemicioglu et al.~\cite{gemicioglu2024breathepulse} explored how airflow delivered by a system can implicitly influence breathing patterns.

\subsection{Diary Study to Understand Situated Use of Sensory Technology}
Among current explorations, one line of work takes a different approach. Rather than introducing new content to support everyday mindfulness, it leverages the rich sensory experiences inherent in everyday interactions. In particular, Tao et al.~\cite{tao2025audio} introduced a system that augments sounds produced through manual interactions, such as rubbing hands together, clinking a cup, or brushing a surface, making sensory cues more salient and helping users anchor their awareness in ongoing activities. In an in-lab study with 60 participants, they found that audio augmentation increased state mindfulness, including both attention and curiosity. Participants noticed auditory details of objects they might otherwise overlook and engaged in trial-and-error exploration, a behavioral indicator of curiosity.

Despite promising results, current evaluations of audio-augmented sensory technology have been limited to controlled, in-lab settings on structured activities. We seek to examine the situated use of the system during everyday mindfulness practices over multiple days. To investigate this question, we chose to conduct a diary study, a qualitative method that collects participants’ in-situ self-reports over an extended period~\cite{rieman1993diary, bastien2010usability, blaynee2016collaborative}. Prior work has leveraged this methodology to observe the experiences and behaviors of small groups of users engaging with emerging technologies, such as mixed reality systems for work and productivity~\cite{cheng2025augmented}, the effects of digital technology on emotional well-being~\cite{smith2022digital, munteanu2026playing}, and assistive devices for deaf and blind users~\cite{herskovitz2023hacking, takaki2023asears}. In mobile health literature, diary studies have been used to inform system iterations and establish feasibility~\cite{bricker2026development, sien2025gentle, kumar2022mcardia}. For instance, Bricker et al. conducted a six-day diary study with seven American Indian and Alaska Native adults to evaluate and refine a culturally adapted smoking cessation app~\cite{bricker2026development}. Sien and McGrenere conducted a week-long diary study with 16 university students to examine how a storytelling platform shaped user engagement with reflecting on mental health experiences~\cite{sien2025gentle}. Diary studies are valuable approaches for understanding situated usage of audio augmentation systems in everyday mindfulness practice out of the lab.

\section{System Implementation}
\begin{figure}[!hbt]
\includegraphics[width=0.6\linewidth]{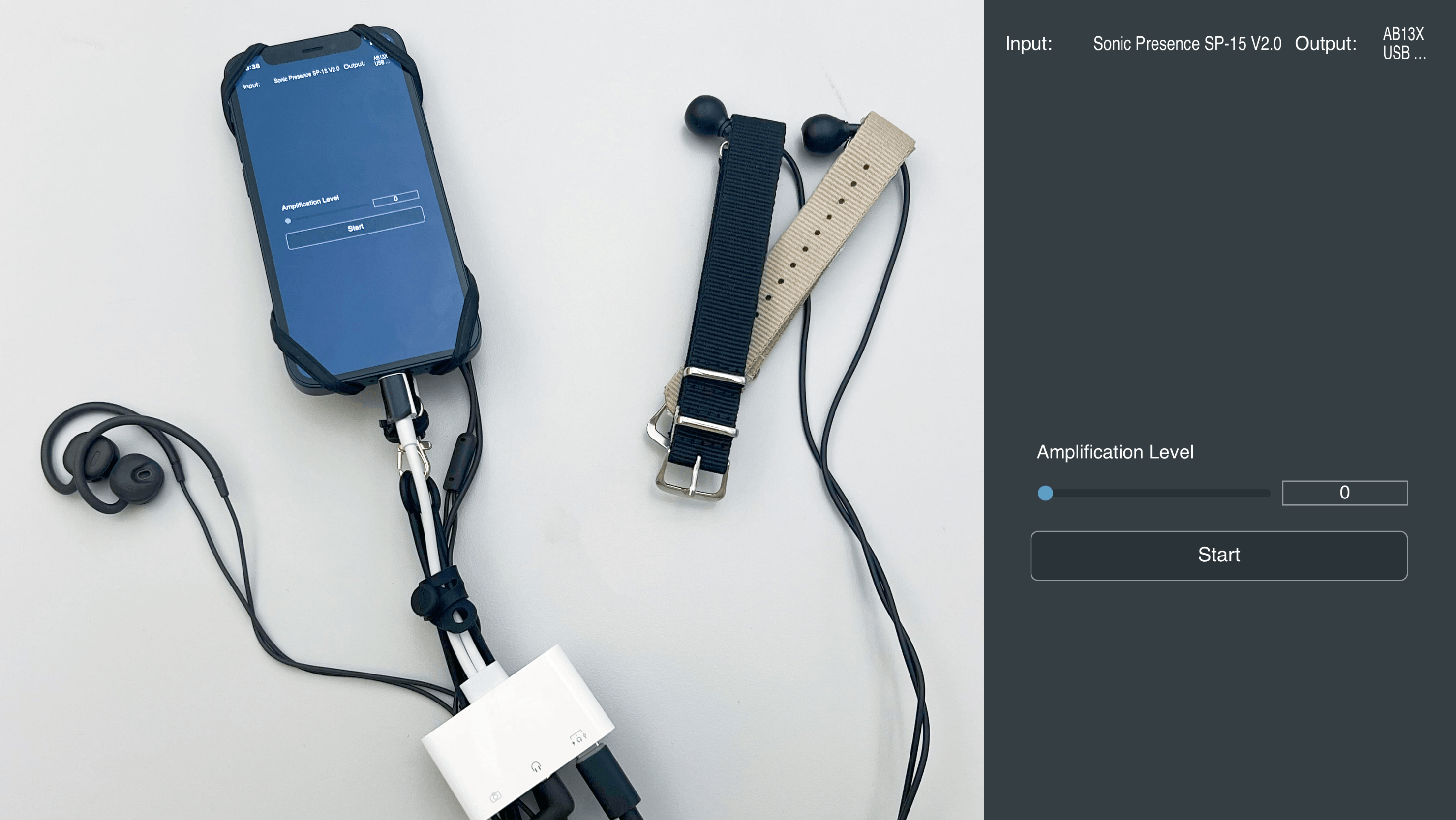}
  \Description{}
  \caption{System implementation. Left: The device kit’s core components—a mobile application, a wristband with microphones, and open-ear headphones. Right: The mobile application interface for adjusting the augmentation level.}
  \label{fig:device_kit}
\end{figure}

To explore how audio augmentation of manual interactions supports everyday mindfulness practice, we needed a system that participants could carry and use with commonly available commercial devices. We therefore developed a wearable system for deployment outside the lab, extending the design introduced by Tao et al.~\cite{tao2025audio}. Figure~\ref{fig:device_kit} illustrates the main components of the resulting device kit.

\textbf{Hardware.} To maintain consistency with the system evaluated in prior work~\cite{tao2025audio}, we adopted the same microphone and headphone hardware, ensuring comparable audio quality and listening experiences. The device consists of a pair of wristbands, each equipped with a microphone (SonicPresence SP15C~\cite{sonicpresence}) to capture sounds generated during users' hand interactions. The amplified sounds are played back through open-ear headphones (nwm Wired Open Ear Stereo Earbuds~\cite{nwm}), allowing users to remain aware of their surroundings while listening to the augmented sounds. Both the microphone inputs and headphone output are connected to an iPhone 12 Mini via an adapter.

To support comfortable use during everyday activities, we further made the system into a wearable setup. The smartphone is carried in a lanyard-style pouch worn around the neck, while all cables are routed behind the user's back. This configuration allows users to move freely, keeps the device readily accessible, and minimizes interference from connecting cables.

\textbf{Software.} The microphone signals are processed in real time through a custom mobile application developed using JUCE~\cite{juce}, a framework for low-latency audio processing. A customized mobile application interface allows users to start and stop audio augmentation and adjust the amplification level by applying digital gain. The application also logs user interactions, including when participants start or stop audio augmentation and adjust the amplification level, enabling researchers to characterize system use patterns.

\textbf{Device Kit.} Each participant received a complete device kit, including an iPhone 12 Mini and a preassembled wearable setup with the wristbands and open-ear headphones integrated into a lanyard. This design minimized setup effort and allowed participants to quickly put on the device and connect it to the smartphone before use.

\section{Study Design}

\subsection{Research Questions}
This work seeks to explore the situated use of audio augmentation of manual interactions in everyday mindfulness practice. Our first research question is: \textbf{RQ1 (Experience):} \textit{How does audio augmentation of manual interactions affect the experience of mindfulness practice during out-of-lab, multi-day use?} The out-of-lab setting also provides a unique opportunity to observe the contexts in which participants naturally incorporate the device into their daily lives. Therefore, our second research question examines: \textbf{RQ2 (Context):} \textit{What contextual factors influence the use of audio augmentation of manual interactions in supporting everyday mindfulness?}  Building on understanding of the user experience and situated use of the device, we further examine how audio augmentation contributes to mindfulness practices: \textbf{RQ3 (Mechanism):} \textit{What role does audio augmentation of manual interactions play in supporting everyday mindfulness practices?} Finally, we seek to inform future iterations of the system by identifying the type of practitioners that would benefit from such a system: \textbf{RQ4 (Users):} \textit{Who would benefit from audio augmentation of manual interactions to support mindfulness?}

\subsection{Procedure}
\begin{figure}[!hbt]
\includegraphics[width=0.8\linewidth]{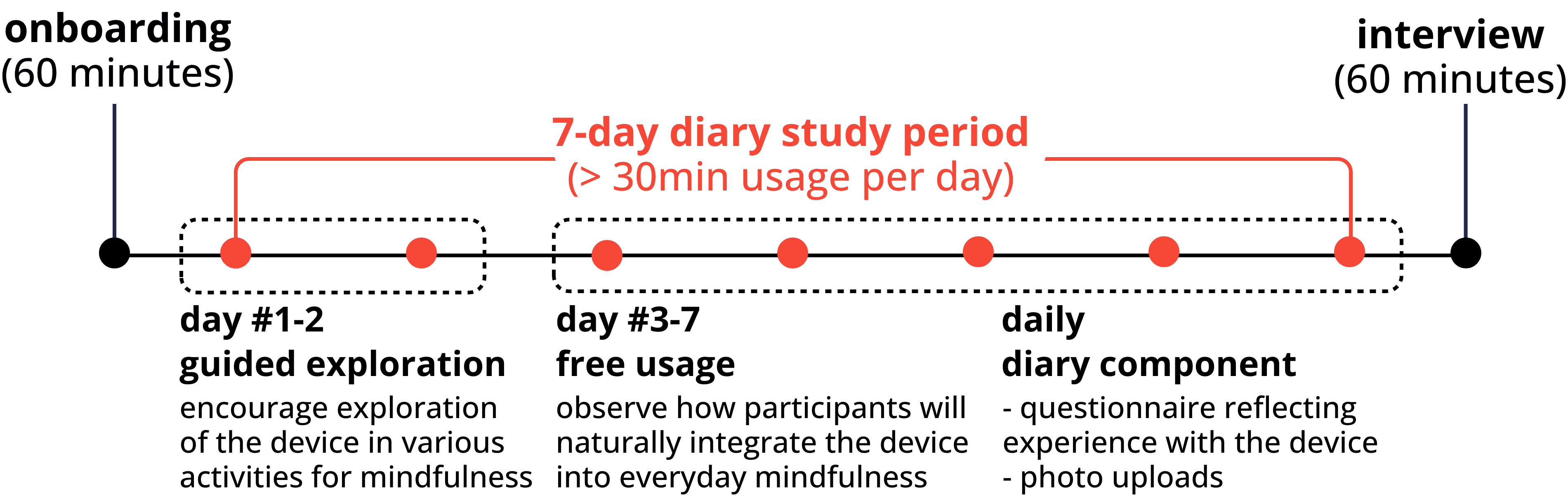}
  \Description{}
  \caption{Study procedure. The study consisted of three phases: (1) an onboarding session, during which participants were introduced to the device and study protocol; (2) a 7-day diary study, in which participants took the device home and used it during everyday activities to support mindfulness. During the first two days, participants completed guided exploration by using the device across different categories of activities. From Days 3–7, participants were free to use the device during any everyday activities of their choice; and (3) a final interview, during which participants reflected on their experiences using the device.}
  \label{fig:study_procedure}
\end{figure}

Figure~\ref{fig:study_procedure} shows the overall study procedure. Participants began with an in-person onboarding session to familiarize themselves with the device. They then took the device home and used it during everyday activities to support mindfulness for a week. The first two days of the diary study were dedicated to guided exploration, and the remaining five days allowed participants to freely choose activities in which to use the device. Participants were asked to use the device for at least 30 minutes each day and complete a diary questionnaire at the end of each day. After the seven-day diary period, participants returned for a final interview.

\subsubsection{\textbf{Onboarding (1h, in-lab)}}
The onboarding session had two goals: (1) introducing the study logistics and (2) familiarizing participants with the system. Researchers first explained the study procedures and the expected daily commitment. Participants were informed that interactions within the device application would be logged. In line with privacy considerations and IRB recommendations, participants were also instructed to use the device only in private spaces, as the system could amplify the voices or other sounds of nearby individuals who are not aware of the device. After reviewing the study procedures and privacy considerations, participants provided informed consent.

During the device demonstration, researchers first showed participants how to wear the devices. Participants were then asked to independently put on the devices and complete several simple activities, such as touching nearby objects and walking around, to become familiar with the system. They also practiced adjusting the audio amplification level using the mobile app. At the end of the onboarding session, participants took the devices home, and researchers provided a handout summarizing the study logistics along with a device manual for future reference.

\subsubsection{\textbf{Device Use (7-day, take-home)}} 
During the diary study, participants were instructed to use the device for at least 30 minutes per day for 7 days. Participants could decide when to use the device, whether to use it in a single session or multiple sessions, and freely adjust the level of amplification. 

\textbf{Day \#1 to \#2: Guided Exploration.} To help participants become familiar with the system and explore the use of the device across different everyday activities, we provided guided instructions during the first two days. The instructions focused on broad categories rather than specific tasks, encouraging participants to try the device in varied everyday contexts and reflect on whether it supported mindful awareness across different activities. On the first day, participants were instructed to use the device in three types of everyday activities: interacting with objects for fun or play, household chores, and information work. On the second day, participants were prompted to use the device in self-care tasks, creative work, and food-related activities. On both days, participants were free to engage in any activities outside of the provided categories, referred to as "Others".

After each day of use, participants reflected on their experiences in a diary questionnaire. The questionnaire asked participants to document the tasks they tried with the device under each type, describe their experiences using the device in these activities, report whether it fostered mindfulness, and note any surprises or challenges they encountered. Participants were also asked to upload at least three photos documenting their device use.

\textbf{Day \#3 to \#7: Free Usage.} In the remaining five days, participants were free to use the device for any activities they preferred. This stage of the study aimed to explore how mindfulness practitioners would naturally integrate the device into their everyday mindfulness practices. No additional constraints were imposed on the number or type of activities completed each day, as long as participants used the device for at least 30 minutes in daily activities with the goal of fostering mindfulness. Similar to the first two days, participants completed a diary questionnaire and photo drop to document their experiences after each day of device usage. 

Researchers sent out daily instructions through email approximately at 6:00 AM each day. If the daily diary questionnaire and photo uploads were not completed, a reminder email was sent after 8:00 PM.

\subsubsection{\textbf{Final-Interview (1h, in-lab)}} 
After completing the 7-day diary study, participants returned the device and participated in a final semi-structured interview. Before each interview, researchers reviewed participants' daily diary entries and uploaded photos to prepare follow-up questions. The interviews consisted of four parts. First, participants were asked to reflect on their overall experience using the device during everyday activities and how it influenced their mindfulness in those activities. Second, we examined the activities and usage patterns in which the device was most effective in supporting mindfulness. This discussion involved reviewing each activity category to identify factors that shaped participants' experiences. Third, we explored the role the system played in participants' everyday mindfulness practices, how it could be integrated into existing mindfulness routines, and what types of mindfulness practitioners might benefit most from it. Finally, participants reflected on challenges they encountered while using the device and suggested potential improvements for future iterations.

% \usepackage{booktabs}
\begin{table}[]
\begin{tabular}{@{}lllllll@{}}
\toprule
\textbf{PID} & \textbf{\begin{tabular}[c]{@{}l@{}}Mindfulness practice \\ experience\end{tabular}} & \textbf{\begin{tabular}[c]{@{}l@{}}Formal mindfulness \\ practice frequency\end{tabular}} & \textbf{\begin{tabular}[c]{@{}l@{}}Everyday mindfulness \\ practice frequency\end{tabular}} & \textbf{\begin{tabular}[c]{@{}l@{}}Coaching \\ experience\end{tabular}} & \textbf{Age} & \textbf{Gender} \\ \midrule
P01 & 3–5 years & multiple times per day & multiple times per day &  & 32 & female \\
P02 & 1-3 years & multiple times per day & multiple times per day &  & 23 & female \\
P03 & 6–10 years & daily & multiple times per day & x & 33 & male \\
P04 & 1–3 years & daily & 3–5 times per week &  & 27 & male \\
P05 & 6–10 years & multiple times per day & daily & x & 32 & female \\
P06 & more than 10 years & 1–2 times per week & 3–5 times per week &  & 36 & female \\
P07 & 3–5 years & 1–2 times per week & 3–5 times per week &  & 31 & male \\
P08 & 6–10 years & 1–2 times per week & multiple times per day &  & 32 & female \\
P09 & 3–5 years & 3–5 times per week & 3–5 times per week &  & 26 & female \\
P10 & 6–10 years & daily & daily &  & 32 & male \\
P11 & 6–10 years & multiple times per day & multiple times per day &  & 24 & female \\
P12 & more than 10 years & multiple times per day & multiple times per day & x & 45 & female \\ \bottomrule
\end{tabular}
\vspace{5px}
\caption{Participant demographics and mindfulness practice experience} 
\label{table:table1}
\end{table}

\subsection{Participants}
In this study, we focused on experienced mindfulness practitioners because their established routines allowed us to examine how the system could fit into existing practice. By reflecting on both their early experiences as beginners and their current practice, experienced practitioners could also offer insights on what type of practitioners can be benefited from the device use.

We recruited participants through local meditation group mailing lists and snowball sampling. The institutional review board approved the study (IRB\# 71992). A pre-screening interest form was used to assess eligibility. Our inclusion criteria were: (1) being at least 18 years old, (2) having established a regular mindfulness practice, (3) having normal hearing without assistive devices, and (4) residing locally so that participants could pick up and return the devices in person. 

Table ~\ref{table:table1} shows the background of our sample. The average age of participants was 31.1 years (SD = 5.8). Eight of the 12 participants self-identified as female, while four self-identified as male. Two participants reported 1–3 years of mindfulness practice experience, three reported 3–5 years, five reported 6–10 years, and two reported more than 10 years of experience. 3 participants have coaching experience for mindfulness and meditation training.

\subsection{Data Collection and Analysis}
\subsubsection{Data Collection}
All participants completed 7 days of device use within 9 days of beginning Day \#1 of the diary study. Seven participants completed the study over 7 consecutive days without interruption, while the remaining five had 1--2 day gaps in device use due to scheduling constraints or technical issues. Final interviews were conducted within two weeks of the initial onboarding session. Due to a technical issue, the device logs for P04’s first day were not saved. However, all questionnaire and interview data were complete for all participants. 

Across the study, we collected three complementary data sources: (1) 84 diary questionnaire entries and 298 photo uploads from 12 participants; (2) device usage logs. Participants contributed a total of 89.84 hours of device usage. On average, participants used the device for 64.9 minutes per day (SD = 58.3); (3) Transcripts of the final interviews reflecting on participants' overall experiences after the diary study.

\begin{table*}[t]
\centering

\label{tab:usage-and-activities}
\small

\begin{minipage}[t]{0.30\linewidth}
\centering
\caption{Device usage in minutes (n = 12 except for day \#1 due to data loss)}
\label{tab:daily-usage}
\begin{tabular}{cccc}
\toprule
Day & $N$ & Mean & SD \\
\midrule
1 & 11 & 53.7 & 26.6 \\
2 & 12 & 76.7 & 62.9 \\
3 & 12 & 69.3 & 82.8 \\
4 & 12 & 76.4 & 76.7 \\
5 & 12 & 68.3 & 49.8 \\
6 & 12 & 50.8 & 37.6 \\
7 & 12 & 58.6 & 58.4 \\

\midrule
Per Day    &       & 64.9 & 58.3 \\
\bottomrule
\end{tabular}
\end{minipage}\hfill
\begin{minipage}[t]{0.34\linewidth}
\centering
\caption{\# of Reported activities for \\ Days~1-2 (n=12 for all days)}
\label{tab:activity-codes}
\begin{tabular}{lr}
\toprule
Category & Count \\
\midrule
Fun or play        & 15 \\
Light chores       & 20 \\
Information work   & 13 \\
Self care          & 22 \\
Creative work      & 16 \\
Food related       & 17 \\
Others          & 24 \\
\midrule
Total            & 127 \\
\bottomrule
\end{tabular}
\end{minipage}\hfill
\begin{minipage}[t]{0.30\linewidth}
\centering
\caption{\# of Reported activities for Days~3-7 (n=12 for all days)}
\label{tab:activities-day3-7}
\begin{tabular}{lr}
\toprule
Day & Count \\
\midrule
3 & 46 \\
4 & 40 \\
5 & 43 \\
6 & 42 \\
7 & 46 \\
\midrule
Total & 217 \\
Avg.\ /participants & 18.1 \\
Avg.\ /part./day & 3.62 \\
\bottomrule
\end{tabular}
\end{minipage}
\end{table*}

\subsubsection{Data Analysis}
We analyzed these data using both qualitative thematic analysis and descriptive statistics. Two researchers independently reviewed the participants' diary responses, uploaded photos, and final interview transcripts using reflexive thematic analysis~\cite{braun2021thematic, clarke2017thematic} to generate initial inductive codes. The researchers then met regularly to discuss their interpretations, compare observations, and iteratively refine the developing codes and themes. The final themes captured recurring patterns regarding the experience of using the system, contexts of use, underlying mechanisms, and potential for integration in everyday mindfulness practice. 

To understand device usage patterns, two researchers extracted activities reported by the participants from the daily questionnaires. Two authors independently coded all reported activities according to their purpose and grouped them into broader categories. Discrepancies were discussed and resolved through consensus. Table~\ref{tab:daily-usage} shows the average usage per day. In the guided exploration phase (Days \#1–\#2), participants reported a total of 127 activity instances, including 15 instances of fun and play, 20 of light chores, 13 of informational work, 22 of self-care, 16 of creative work, and 17 of food-related activities. Participants also reported 24 activity instances outside the six designated exploration categories. On average, each participant engaged in 1.4 activities per designated category and completed an additional 1 activity outside these categories per day. In the free exploration days (Day \#3 to Day \#7), a total of 217 activities are reported by 12 participants. On average, participants reported doing 3.6 activities with the device per day. A summary of the reported activities is detailed in Table~\ref{table:table2}.

To understand how participants’ experiences with the device changed over time, we aggregated their experience descriptions into 84 entries, with one entry per participant per day. We analyzed each entry using the NLTK VADER sentiment analyzer~\cite{hutto2014vader}. We then computed summary statistics of sentiment scores for each participant and each day, as later discussed in Figure~\ref{fig:sentiment_analysis}.

\section{Findings}
In this section, we present findings from the thematic analysis alongside descriptive statistics. Each main finding addresses a research question, while the subsection headings identify the themes developed through our analysis.

\subsection{Finding \#1: Practitioners Found Audio Augmentation to Support Their Mindfulness Practice Across Multiple Days}

\begin{figure}[!hbt]
\includegraphics[width=\linewidth]{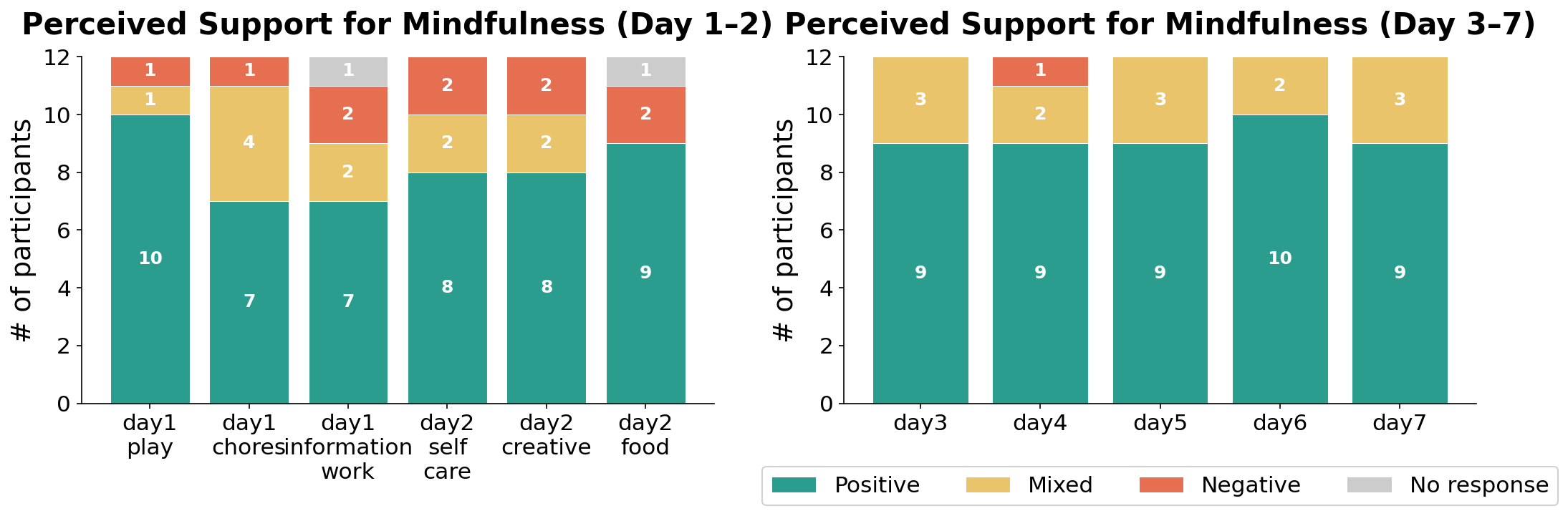}
  \vspace{-20pt}
  \caption{Responses distribution in the diary questionnaire on whether the audio augmentation system supported mindful engagement. Responses were coded into four categories: \textit{Positive} indicated that participants explicitly reported increased mindfulness due to the system. \textit{Negative} indicated that participants reported no benefit or that the system actively detracted from their experience. \textit{Mixed} responses reflected both positive and negative effects, or cases where the impact depended on the context of use. \textit{No Response} was assigned when participants left the question unanswered.}
  \Description{}
  \label{fig:mindfulness_helpful}
\end{figure}

To understand how augmentation of manual interaction affects the experiences of everyday mindfulness practices, we first analyzed diary entries in which participants reflected perceived support from the system. We coded responses based on the nature of the reported experience: solely positive (green), mixed positive and negative/no effect (yellow), solely negative or no effect (red) or no response to how the device impacted on their experience (grey). Figure~\ref{fig:mindfulness_helpful} shows the distribution of the coded responses from participants. Among 132 diary responses, 95 responses (72\%) indicated a positive impact. 

With mindfulness being a multifaceted construct, we further examined the diary questionnaire and interview accounts based on its two components: (1) self-regulation of attention toward the present moment, and (2) an attitude of curiosity, openness, and acceptance toward that experience~\cite{bishop2004mindfulness}. Consistent with the laboratory findings of Tao et al.~\cite{tao2025audio}, participants reported how audio augmentation impacts both components of mindfulness during everyday activities. In 5.1.1 and 5.1.2, we reported findings by each component of mindfulness.

% crinkle of the bag I was holding (P07)
\subsubsection{\textbf{Audio Augmentation Enhances Awareness of Everyday Interactions}} Everyday interactions are rich with sensory details. By amplifying auditory details naturally present within these interactions, one way the device facilitated mindfulness is by directing participants' attention toward the activities themselves. This included bringing familiar sounds back into awareness. For instance, participants described becoming more aware of sounds such as \textit{``opening/closing pens with caps''} (P09) \textit{``creaking of the door handle''} (P11) and \textit{``water pouring from the tap''} (P10). The device also enhanced awareness of subtler details and P07 described them as \textit{``hand movements I normally make [but] had never heard the sound before''}. This could be \textit{``sound of the pen meeting the paper and the strokes of drawing my own signature''} (P05), \textit{``sound of different blooms brushing up against each other''} while snipping flowers (P12) and the sound of \textit{``a piece of clothing falling onto the bed''} while folding laundry (P03).

The amplified auditory details brought greater awareness to the activities and objects involved in the interaction. P02 described their experience of holding a book: \textit{``I hear the sound of my fingers against the page and the page flipping sound very clearly. It was very satisfying (and something e-books can't give). So this made me more aware of the physical presence of the objects I interact with.''} Similarly, P01 shared how an everyday activity like \textit{``peeling an orange''} felt richer:

\begin{quote}
~\textit{``...it's such a regular activity that we do, but just by having all of those little amplifications of different sounds that you don't pay attention on...it just makes everything more deep. You know that you uncover different layers in every activity that you do. So that's why it's enjoyable.''}
 \end{quote}
 
Such heightened awareness also extended to activities otherwise performed automatically. As P03 described:

\begin{quote} 
~\textit{``Replacing trash bags is usually not an activity that I perform with a lot of awareness, but the act of pulling out the trash bag, tying it, taking out a new trash bag, opening it (which is quite a vibrant and exciting experience when you are wearing the device), and fitting it on the trash can were all more drawn into my awareness.''}
 \end{quote}

\subsubsection{\textbf{Audio Augmentation Sparks Curiosity in Mundane Activities}} Beyond the attentional aspect, mindfulness is also characterized by an attitude of openness and curiosity toward present-moment experiences~\cite{bishop2004mindfulness}. Participants reported how hearing everyday interactions in a new way sparks curiosity in mundane activities. For instance, P04 shared their experience of opening the package and eating a chocolate bar: ~\textit{``It was very compelling. The sounds felt unexpected because this kind of interaction is not usually very salient from an auditory perspective. That made me curious and attentive to the subtleties.''} Similarly, P01 described:

\begin{quote} 
~\textit{``And it also made me more curious about the things that I was doing, because a lot of the noises that were amplified, you can't really recognize them because they're too low. But when they're amplified, you're curious about, you know. You want to know what's going on. Why does this sound like this?''}
 \end{quote}

Curiosity, in particular, can manifest in user behavior, such as developing more exploratory patterns. We found that participants reported changes in how they engaged with mundane activities as a result of the audio augmentation. For instance, they shared that they would \textit{``listen to sounds a lot more''} (P10), and \textit{``kept me wanting to interact with more pens to see what other sounds they make''} (P09). P05 described how they approached organizing a small handbag differently:

\begin{quote}
~\textit{``I actually slowed down and started picking up each object a little bit more mindfully. I was holding the keychain, the keys, the cone, the pens in my hands with more gentleness as I placed them neatly into the bag in an orderly manner.''}
 \end{quote}

Similarly, P10 reflected on how using the device changed the way they eat: ~\textit{``...it was much easier to appreciate the sounds, that come from eating, and also, I'm gonna admit sometimes when I eat, you know, I put on a video, or I put on some music or something to help, but with this (device), I just stayed silent throughout, and I gained a new appreciation as a result, I really learned to, savor each bite.''}

\subsubsection{\textbf{Perceived Support to Mindfulness Practices Persists over Multiple Days}} 
\begin{figure}[!hbt]
\includegraphics[width=\linewidth]{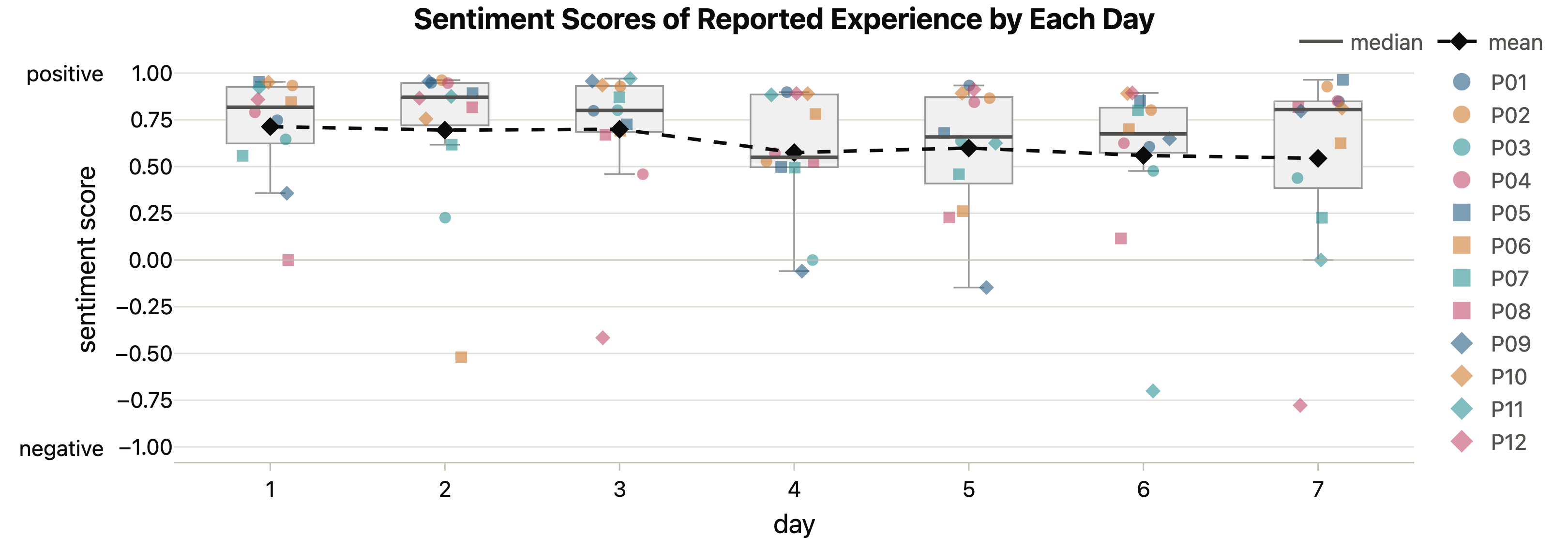}
  \vspace{-20pt}
  \caption{Sentiment scores of participants' reported experiences in the diary questionnaire by study day and participant. Sentiment scores range from -1 (most negative valence) to 1 (most positive valence). On average, participants' descriptions of their experiences with the device exhibited highly positive sentiment.}
  \Description{}
  \vspace{-5pt}
  \label{fig:sentiment_analysis}
\end{figure}

Understanding how the impact of audio augmentation of manual interaction evolved over multiple days is core to our research question. To explore this question, we first conducted a sentiment analysis examining the emotional valence associated with participants' reported experiences of using the device. We aggregated all experience descriptions into 84 entries (one entry per participant per day) and analyzed each entry using the NLTK VADER sentiment analyzer~\cite{hutto2014vader}. This process yielded sentiment scores ranging from -1 (most negative valence) to 1 (most positive valence). Figure~\ref{fig:sentiment_analysis} shows the distribution of sentiment scores across the seven days of the diary study.

We observed that overall sentiment remained consistently positive throughout the 7-day usage period, with Day \#1 ($M = 0.71, SD = 0.29$), Day \#2 ($M = 0.70, SD = 0.44$), and Day \#3 ($M = 0.70, SD = 0.38$) all exhibiting average sentiment scores above 0.7 (i.e., highly positive). The biggest fluctuation in sentiment happened at Day \#4 ($M = 0.58, SD = 0.33$), after which sentiment remained relatively stable on Day \#5 ($M = 0.60, SD = 0.34$), Day \#6 ($M = 0.56, SD = 0.45$), and Day \#7 ($M = 0.54, SD = 0.51$). This trend is consistent with a potential novelty effect, in which initial exposure to the device elicits higher enthusiasm that gradually stabilizes as participants become familiar with its use. However, the persistently moderate to high levels of positive sentiment suggest that participants continued to experience favorable interactions with the system over the 7 day period.

Qualitatively, we observed a similar trend. Participants acknowledged that the novelty exists when first experienced audio augmentation of their manual interaction sounds. Although the novelty wore off over time, its impact on mindfulness continued to be positive. For instance, P02 described: \textit{“The sense of wonder and curiosity peaks in the beginning just because of the novelty, and then it does decrease over time. I think that's the general pattern, but it still helps me focus pretty well.”} P11 similarly noted: \textit{“The audio input is increasingly normalized so the novelty is wearing off but I still think it's useful.”} P05 suggested that the fading of novelty may itself support a calmer mental state to fully leverage the power of the device:

\begin{quote}
\textit{``The first couple days, I was just kind of doing it, and I was interested, I was curious, but not but there was still some wandering in my mind. There's still some thoughts. But then by the third day or fourth day, I was excited to wear the device and be like, Ah, this is my mindful time. I don't have anything else I have to do except for what's right in front of me.''}
\end{quote}

Repetitive use of the device in the same activity could lead to \textit{``desensitization''} as P07 described, but could also increase awareness of previously “boring” tasks, as P08 shared:

\begin{quote}
\textit{``I repeated a lot of the activities, the ‘boring’ ones especially. And I found a lot of joy. I guess my point is the boring ones aren’t boring at all if I am present enough to be with them. Repetition with presence feels completely different from repetition on autopilot.''}
\end{quote}

P03 similarly argued that even with the same task, the exact interaction is not always the same: \textit{``although each time I'm approaching an activity, including with the device, it's going to be a different experience. So in that way, like each day using it was different, and then my state of mind, right, is going to be different each time I'm using the device.''}

\subsection{Finding \#2: Sound Characteristics and Task Demand Influence the Utility of Audio Augmentation}

As shown in Figure~\ref{fig:mindfulness_helpful}, while audio augmentation of manual interactions was generally perceived as supportive of mindfulness, participants also reported more nuanced experiences. In 24 responses (18.2\%), participants indicated that although the device was beneficial, its effects were not uniformly positive across all contexts. An additional 11 responses (8.3\%) described either no effect or a negative effect on mindful engagement. Through an analysis of these responses alongside interview accounts, we identified several factors that shaped the suitability of audio augmentation to support mindfulness in everyday activities.

\subsubsection{\textbf{Characteristics of Natural Sounds Shape Augmented Sensory Experience}}

10 out of 12 participants suggested that the acoustic characteristics of activities affect the suitability of audio augmentation. 

\textbf{Auditory Saliency.} On one hand, when certain sounds are already salient, further amplification may become counterproductive. For instance, participants noted examples such as sound from \textit{``running water from a sink''} (P04), \textit{``when dishes clash each other''} during cleaning (P06), or \textit{``gushing wind sound''} on a bike ride (P02) to be already salient. Instead of drawing attention or sparking curiosity, amplifying already salient sounds could be deemed as "jarring" (P11) or "distracted my attention" (P09). Reflecting on the role of a sound's inherent characteristics, P06 contrasted different activities:~\textit{``Washing hands and brushing was not so good since it got too loud even with the adjustment. Flossing was good since it made consistent soft sound.''}

P03 shared a story of them deciding whether they should use the device while vacuuming: \textit{``I actually thought about this, like, should I use it while vacuuming? ...I'm like, no, I don't think I really need to do that.''} Furthermore, P03 elaborated that:

\begin{quote}
\textit{
``...for things we're already pretty attuned to the sound, I think we probably don't need to use the device as much to remind us.''}
\end{quote}

On the other hand, when hand interaction barely produces sounds, participants also found augmenting the sounds to be less helpful. P06 described their experiences of embroidery \textit{
``Since a needle is used, embroidery actually did not make a lot of sound.. Because of lack of activity sound...there are some activities that might not get a lot of help from sound.''}

\textbf{Auditory Predictability.}
Augmenting natural sounds in everyday interactions further introduces the challenge that salient sounds may occur unexpectedly, making it more difficult for users to maintain mindful awareness. P12, for instance, described one such experience with the device:~\textit{``I was petting my cat, and it was really lovely, until she meowed right into the microphone, or I was outside snipping roses and, like, that was really cool, until dogs started barking, and then it's, like, way too loud in my ear.''} Similarly, P02 shared: \textit{``One scenario that I really don't like using it is, for example, when I'm cooking, everything's going great, but then my knife hits the plate, and there's this just loud, like, boom in my earphone.''} Therefore, P09 suggested that the device may be best suited for \textit{``activities that I know what to expect, kind of like what sounds are gonna come and they're like, not loud.''} While any stimuli could have been anchors for grounding mindful attention, P11 emphasized that loud and abrupt sounds make the process harder:

\begin{quote}
\textit{
``And I think I can use any sound to bring myself back to the moment, but it's just more pleasant when it's a soothing or enjoyable sound.''}
\end{quote}

To mitigate this issue, P06 suggested that future iterations of the system \textit{``control the volume automatically''} when loud and abrupt sounds occur.

\begin{table}[]
\begin{tabular}{|ll|ll|ll|ll|}
\hline
\textbf{Housework}  & \textbf{54} & \textbf{Information Work}     & \textbf{41} & \textbf{Personal Care \& Hygiene} & \textbf{25} & \textbf{Transit}       & \textbf{13} \\
preparing food      & 12          & working                       & 19          & brushing teeth                    & 13          & walking                & 8           \\
cleaning dishes     & 9           & reading                       & 8           & doing skincare routine            & 4           & driving                & 4           \\
doing laundry       & 6           & typing                        & 7           & washing hands                     & 3           & biking                 & 1           \\
organizing clothes  & 6           & emailing                     & 2           & personal care                     & 2           & \textbf{}              & \textbf{}   \\ \cline{7-8} 
preparing beverage  & 6           & writing                       & 2           & combing hair                      & 1           & \textbf{Creative work} & 5           \\
feeding pet         & 3           & journaling                    & 1           & flossing                          & 1           & painting               & 3           \\
cleaning the floor  & 2           & using a computer              & 1           & taking vitamins                   & 1           & assembling gift box    & 1           \\ \cline{5-6}
managing trash      & 2           & virtual meeting               & 1           & \textbf{Leisure}                  & \textbf{16} & designing              & 1           \\
plant care          & 2           & \textbf{}                     & \textbf{}   & socializing                       & 4           &                        &             \\ \cline{3-4}
ironing clothes     & 1           & \textbf{Eating \& Drinking}   & \textbf{37} & playing cards                     & 2           &                        &             \\
opening door        & 1           & eating                        & 27          & texting                           & 2           &                        &             \\
opening packages    & 1           & drinking                      & 10          & laying down                       & 1           &                        &             \\
organizing items    & 1           &                               &             & listening to music                & 1           &                        &             \\ \cline{3-4}
purchasing food     & 1           & \textbf{Meditative Practices} & \textbf{26} & looking out of the window         & 1           &                        &             \\
putting things away & 1           & meditating                    & 14          & playing with objects              & 1           &                        &             \\
                    &             & yoga                          & 6           & ring bell                         & 1           &                        &             \\
                    &             & stretching                    & 3           & sleeping                          & 1           &                        &             \\
                    &             & exercising                    & 2           & talking                           & 1           &                        &             \\
                    &             & chanting                      & 1           & watching videos                   & 1           &                        &             \\ \hline
\end{tabular}
\caption{Activities in which participants used the device during Days 3–7 of the diary study. The table summarizes the frequency of individual activities and their corresponding broader activity categories. Eating was the most frequently reported individual activity, while housework was the most frequently reported broader activity category.} 
\label{table:table2}
\vspace{-10px}
\end{table}

\subsubsection{\textbf{Task Demands Influence Available Bandwidth to Engage with Audio Augmentation}}

Equally, 10 out of the 12 participants noted that the cognitive demands of an activity influence its suitability for audio augmentation. In particular, when an activity is already cognitively demanding, there may be insufficient attentional bandwidth to engage with augmented auditory experiences. 

\textbf{Familiarity and Repetitiveness Reduce Task Demand}. An activity could be perceived as low in demand due to its repetitive and familiar nature. P01, for instance, included \textit{``folding the laundry''} and \textit{``skin care routine''} under this category. P06 similarly mentioned \textit{``typing''}, \textit{``eating, even chewing sounds''} as repetitive activities that could benefit from audio augmentation. P09 described food preparation as their familiar routine, noting that: \textit{"I know what's coming. I know that I'm gonna break this egg. I know how much cheese I'm gonna put like, I know all the steps. So I was able to feel familiar with those sounds"}. 

These qualitative accounts align with the activity frequencies identified from participants' diary entries, as shown in Table~\ref{table:table2}. Housework activities were the most frequently reported category, occurring 54 times during the 5 days free usage and encompassing activities ranging from food preparation to plant care. Eating was the single most frequent activity, occurring 27 times between Day \#3 to Day \#7.

When activities are low in cognitive demand, participants suggested that opportunities emerge to direct additional attentional resources toward deeper engagement with sensory experience. \textit{``I think that when something is repetitive, it kind of loses its meaning. You know, you just do it because it's repetitive, but if you have a device that helps you understand more deeply what's going on, the experience just takes a different perspective,''} said P01. Similarly, P02 echoed:

\begin{quote}
\textit{
``When I'm doing something that's less cognitively demanding and that I am more prepared to feel a sense of wonder, it helps in that direction.''}
\end{quote}

\textbf{Task Demands Vary Across Contexts and Individuals.} However, there is no clear-cut distinction between activities that are low in cognitive demand based solely on their functional intent, as this is often shaped by contextual factors. For instance, unlike P09, P01 described meal preparation as unsuitable for audio amplification due to time constraints they often experience: \textit{``...because I had a limited time span to do that, I wasn't really paying attention to the sounds.''} P02 further emphasized that the relationship between the task and the individual also matters:

\begin{quote}
\textit{
``I think it matters how all of these tasks fit into a person's life... like someone's whole job might be cooking and they are a Michelin chef and it's a lot of cognitive work, or someone like me who just needs to eat and my main job is doing research... It's not just about the task, but how this task relates to the person doing it.''}
\end{quote}

Therefore, for informational work activities—which are themselves heavily cognitive rather than physical—we observe a split in participants' attitudes regarding whether the device is beneficial, depending on the demand of the task. Participants who found the device to be helpful described the work they did as \textit{
``house keeping''} (P09) or \textit{``administrative inputting stuff''} (P12) rather than \textit{``deep thinking''}. P06 described: \textit{``checking emails, that's like, really quick. I didn't really plan it for using [the device], for example, lab work or like experiments.''} \textit{``I think for my deepest thinking, I usually like to just have silence''} echoed by P11. P07 expressed a more extreme view regarding highly cognitive activities:\textit{``Is it even possible to be mindful of reading? And I think, no, like, because I have to stop being aware of my sensory experience and just be fully, you know, deeply engaged intellectually with this thing, and the same thing with writing emails or doing homework or something.''} 

\subsubsection{\textbf{Additional Factors}} Beyond sound characteristics and task demands participants identified several additional factors influencing the utility of audio augmentation in supporting everyday mindfulness. 

\textbf{Meaningful Hand Interaction.} One such factor is whether the activity involves meaningful hand interaction. For instance, P11 described a case in which audio augmentation was of limited benefit due to the lack of hand movement in the activity they performed: ~\textit{``Because my hands were mostly idle during the more mindful experiences I did (meditation) and I was in an already quiet place, it didn’t had much of an impact on enhancing or detracting from my mindfulness.''}

\textbf{Intention for Mindfulness.} Another factor influencing the utility of audio augmentation lies in users’ intention to practice mindfulness. Both P07 and P08 noted occasions in which they did not intend to engage in mindfulness and instead rushed through their tasks. In such cases, even when sounds were amplified, the augmentation did not foster mindfulness as effectively as it might when they had the intention to engage mindfully. P07 described wearing the device during their morning routine and reflected on whether the audio augmentation helped:~\textit{``It did, but I wouldn’t say my mindfulness felt strongly sustained, especially across tasks. I think part of it has to do with the fact that I have very much trained myself to do these activities as fast as possible so I don’t have a habit of mindfulness during these activities.''}

\textbf{System Form Factor.} Last but not least, the form factor of the current system occasionally constrained the utility of audio augmentation. Participants reported that the wired device setup was inconvenient in certain contexts, particularly during activities involving water (P12, P09), which in turn could distract from mindful attention.

\subsection{Finding \#3: Audio Augmentation Scaffolds Mindfulness but Does Not Replace Deliberate Practice}

Building upon our findings regarding how audio augmentation affects mindfulness and the factors that shape its utility, we next examine the role that augmenting natural sounds plays in the mindfulness process. Understanding the mechanisms through which audio augmentation supports mindfulness can provide insight into how such systems may complement existing mindfulness practices and training approaches. Participants' reflections suggested several distinct but related mechanisms.

\subsubsection{\textbf{Capture Attention into Present Moment Interactions}} One prominent mechanism that emerged from participants’ interview accounts was how audio augmentation pulls attention away from a wandering state and toward present-moment sensory experience. By augmenting sounds from ongoing interactions, the device “captures” (P02) users’ attention, meaning that the shift is primarily bottom-up and sensory-driven. 

Participants reported that this attentional capture was especially useful during transitions into mindfulness, which could otherwise be challenging when they were \textit{"super busy with other things"} (P06), when they were \textit{"so swamped up in life that I forget"} (P02), or when they had \textit{"got a little stressed"} (P11). P04 described one such experience of transitioning from a long day of work into mindfulness practice: 

\begin{quote}
\textit{“When I meditate in the evening, I'm transitioning from being in conversations, from doing things the entire day, right? And my mind is full of thoughts and ideas...and I start meditating, I almost sort of like force myself to meditate because I know it's going to be good for me...But that process of me sort of actually surrendering to the meditation itself takes some time.”} 
\end{quote}

Using the device, P04 instead described this transition as having "less friction" and said:

\begin{quote}
\textit{
"...it almost felt like I automatically was in that state, because I'm listening by default a lot more to all of these sounds that are in my environment and that only happens after a few minutes in the meditation when I'm like, Oh my God, there's actually so many things happening around me and and I wasn't realizing, because I was thinking about what I'm going to do tomorrow."}
\end{quote}

Similarly, P01 described how audio augmentation made cues from the present moment interactions more salient as compared to the internal thoughts:

\begin{quote}
\textit{
``Other thoughts are kind of blocked...because you have more ways of occupying your attention and so your attention instead of going to somewhere outside...your attention is more focused on what you're doing right now, because you have more stimuli that are coming in from wearing the device.''}
\end{quote}

\subsubsection{\textbf{Embodied Anchors to Sustain Mindfulness}} Beyond the initial capture of attention to transition into a mindful state, participants also emphasized how the audio augmentation system helps sustain their mindfulness due to its embodied nature. 

P05 and P11 contrasted the augmentation of natural sounds with existing mindfulness technologies for sustaining mindfulness. Existing tools rely on explicit instructions and pre-scripted visual or auditory stimuli and P05 described such approaches as attempting to \textit{``control the mind with the mind.''} While they acknowledged that these methods can be effective---\textit{``sometimes it works if you're really willing to follow those instructions''}---they also noted that it can be difficult to sustain mindfulness with only instructions, because \textit{``the mind is a little bit like an untrained puppy that...just wants to run all over the place.''} P05 pointed out how the audio augmentation system instead operates through a different mechanism:\textit{
"...instead of trying to change their thoughts, you're just giving them something else to pay attention to."}

P11 further emphasized that the anchors provided by the system were fundamentally different from pre-recorded soundscapes such as ocean waves because they were directly tied to the user's own actions. They highlighted the \textit{``physical control over the sound''} as a key characteristic of the system:

\begin{quote}
\textit{
``Listening to just like [pre-recorded] nature sounds on my phone is a little janky sometimes or feels artificial. So I think [the device] is like maybe providing another way to access that mindfulness modality in a way that is relaxing and satisfying, and also like physically embodied,''} said P11. 
\end{quote}

\subsubsection{\textbf{Experiential Target for Mindfulness Practice}}
Beyond facilitating entry into mindfulness and sustaining mindful states, several participants also suggested that augmenting ecological sounds could provide an experiential target for practitioners to work toward. In this sense, the enhanced sounds of manual interactions may provide a concrete sense of what it feels like to be deeply aware of everyday activities. P07 elaborated:

\begin{quote}
\textit{``It can be really helpful if you've never really quite had the experience of or understand what that feels like... it gives you, like, a sense of what it could feel like to be very aware of the sound of your hands, you know.''}
\end{quote}

P04 argued that such referential experiences could encourage continued practice:

\begin{quote}
\textit{
``It's like you're facilitating the entry level to a lot of people so that they have that reference point that allows them to know that there is something they can attain by committing to the practice, and that's huge. I think that has a lot of value. It's like an entry point to knowing, to attaining that experience that will allow you to justify the practice in a way that will give you that reference that this thing is attainable and it's real, and therefore you will put the work together without the device.''}
\end{quote}

Furthermore, P03 suggested that real-world sounds themselves, being augmented before, could continue to support practice beyond device use:

\begin{quote}
\textit{
``...it's a good reminder when I'm not wearing the device to practice and to sort of look for those sounds that you know maybe I won't hear as clearly, but trying to just notice, what can be there?''}
\end{quote}

\subsubsection{\textbf{Does Not Replace Deliberate Practice}}
As much as the device could scaffold the users into mindfulness, participants also emphasized that it did not replace deliberate practice. Rather than creating mindfulness automatically, participants noted that mindfulness still required an active intention to attend to the current moment for the device to be beneficial. P08 articulated this distinction clearly:

\begin{quote}
\textit{``I was trying to rush through the experience of mundane things and was not present. But then when I try to be present, the device helped a lot.''}
\end{quote}

P08 made a metaphor comparing the experience of fostering mindfulness to going to the gym, where practitioners need to \textit{``...go through the journey of watching your mind being occupied with thoughts, watching your mind running wild and crazy and being able to breathe and be patient with it.''} Similarly, P11 emphasized~\textit{``the most positive impact of mindfulness is actually through practicing it.''} Therefore, without the intention to practice mindfulness, wearing the device does not automatically lead to enhanced awareness. As P07 explained, ~\textit{``because, you know, you could have the device on...and then just kind of like after the first few minutes...oh, I forgot I'm even wearing this thing.''}

In the long term, practitioners ultimately want to stay mindful on their own. P01 raised the potential risk of becoming dependent on the device, noting that there may be a possibility that \textit{``the only way you can be mindful of specific activities is when you have the device on.''} P02 addressed this concern by suggesting a design that \textit{``starts with a high volume, but then it kind of decreases on its own and guides the person to do it.''} Similarly, P03 proposed that the role of the device should be \textit{a ``tap on the shoulder,''} and P02 nicely summarized this perspective:
\begin{quote}
\textit{
"You need something to pull people into that state, and once you flow into that state, ultimately you want someone to be able to cultivate on their own and repeat it again and again, without a specific device, I think, ultimately, we want people to be more mindful in everything."}
\end{quote}

\subsection{Finding \#4: Both Novice and Experienced Practitioners Could Benefit from Audio Augmentation}

Studying experienced mindfulness practitioners allowed us to explore which types of practitioners might benefit from the device. Participants could draw on both their experiences learning mindfulness and establishing a  practice routine. Three of our participants had also served as coaches in mindfulness classes. Overall, 9 participants perceived the device to be beneficial for beginner practitioners and 5 participants perceived the system suitable for experienced practitioners, while the device could serve different roles at different stages of experience level.

\subsubsection{\textbf{Lower the Threshold for Mindfulness Practice Among Novices}}
Novice mindfulness practitioners often face challenges entering mindful states during everyday activities, which can reduce their motivation to practice consistently. For instance, P03 described that novices may \textit{``haven’t really intentionally done that sort of mindful awareness of everyday tasks''}. Even when they attempt to practice, beginners may \textit{``don’t see improvements”} (P01), making it difficult to recognize the benefits of mindfulness and sustain long-term engagement. 

With augmentation of ongoing interaction sounds, P08 described that the system helps beginners to\textit{``cut that journey''} and \textit{``helps to get into a state of being mindful a little bit faster''}: 

\begin{quote}
\textit{
``It instantly picks you up and drops you in an environment where you have to pay attention to the sound, it's so vivid. And I mean, you open your eyes, you wash your hands right in front of you, the image is vivid in front of your eyes, and now it's being amplified with the sound.''}
\end{quote}

Lowering the barrier to entering a mindful state may broaden the accessibility of mindfulness practice for individuals who otherwise \textit{``never think about mindfulness''} (P09). P12 further elaborated:

\begin{quote}
\textit{
``A lot of people can have this experience of noticing their environment and the things they do without necessarily having to do a lot of mindfulness and a lot of really, really heavy lifting training themselves to get to that point where they can listen to the environment. So I think, yeah, I really like it, because I think it can facilitate that experience that would incentivize the practice itself.''}
\end{quote}

In terms of integrating the system into novices' mindfulness routines, P11 suggested that beginners may require more structured guidance than simply being given the device, as participants were in this diary study. Participants proposed strategies for scaffolding novice use. In particular, P05 recommended novices to start with wearing the device during activities they enjoy:~\textit{``I think it's a good idea to start with activities they already enjoy, and wear the device and see if it amplifies their joy.''} P05 further suggested using the device continuously for 15--20 minutes at a time, allowing users to settle into and deepen their engagement with the practice. 7 out of 12 participants also cautioned that frequently turning the amplification on and off may interrupt the development of sustained attention and encourage repeated transitions in and out of mindful awareness.

\subsubsection{\textbf{Support Continued Deepening of Mindfulness Practice for Experienced Practitioners}}
Five participants discussed how the audio augmentation system could benefit experienced practitioners who already have established mindfulness practices.  Participants emphasized that mindfulness is a lifelong journey and that, in some sense, \textit{``everyone is a beginner in mindfulness,''} as P03 remarked. Similarly, P11 highlighted that the device, as a "mindfulness enhancing tool" could benefit users \textit{``regardless of mindfulness experience, because it helps you come back to the moment by enhancing the sounds of, what your hands are doing in that moment.''}

Unlike beginners, experienced practitioners require less structure in integrating the device into their mindfulness routines. Instead, they may benefit more from opportunities to explore how the device fits within their existing practices. As P11 explained, \textit{``it's maybe more just like getting acquainted with the device and people finding what activities or like situations that works best for them in''}, said P11. Similarly, P01 suggested using the device \textit{``in cases where you know you just need to focus a lot, and you put the device on so you're doing work and you focus.''} These accounts suggest that experienced practitioners may appropriate the device in a more self-directed manner, tailoring its use to specific contexts and attentional goals.

P05 suggested that the progression of practice with the device may differ between novice and experienced practitioners. While novices might begin by using the device during activities they already enjoy, experienced practitioners could instead engage with it during activities that are \textit{``actually uncomfortable or boring and see if even those activities that are uncomfortable and boring can bring you joy.''} P05 further elaborated:

\begin{quote}
\textit{
``Like for me, I intentionally ate boring food to see if I like it; and I intentionally did some chores and tasks, like trainings and things like that, modules that I would find boring to see. Does this enhance my mindfulness? And in many cases, it did. Oh, interesting.''}
\end{quote}

\section{Discussion}

\subsection{Situated Use of Audio Augmentation in Everyday Mindfulness Practice}
This work presents the first out-of-lab investigation of audio-augmented sensory technology in everyday mindfulness practice. The week-long diary study with 12 experienced practitioners offers insights on how augmenting manual interaction sounds impacts practice experience, when it can support mindfulness practice and what factors might limit its utility.

Firstly, we observed encouraging evidence for using audio augmentation to support everyday mindfulness: experienced mindfulness practitioners found a positive impact of the device on their mindfulness routine over multiple days. Participants highlighted a distinctive characteristic of the system: the sounds surfaced by the system arose directly from their own actions. Such coupling enabled the system to provide embodied anchors for grounding awareness rather than splitting attention between task and the mindfulness guidance. These findings suggest that everyday mindfulness places different demands on technology than formal practices such as sitting meditation, where practitioners set aside time and space to follow guidance. Systems for formal practice often introduce stimuli such as virtual soundscapes~\cite{cochrane2020mediscape}, haptic feedback~\cite{dauden2021interoceptive}, or visual animations~\cite{liu2022mindful}. However, in everyday mindfulness, when users are already engaged in an everyday activity, tying support to their actions may better support users to weave mindfulness into what they are doing.

Despite these benefits, supporting mindfulness by augmenting naturally occurring sounds means that the device experience depends in part on the acoustic characteristics of users’ interactions. For instance, one participant reported that using the system while patting a cat was satisfying until it unexpectedly meowed. These findings align with prior work suggesting that bottom-up, sensory-driven auditory attention is captured by salient events~\cite{huang2020push}. In this case, the device could accidentally make unintended events more salient, and the unexpected sounds could interrupt focus. Moreover, participants reported that the natural salience of surrounding sounds also played a role: amplifying a sound that already commands attention may only be jarring, and it would be more beneficial to bring subtle, otherwise overlooked sounds into awareness. These findings suggest that audio augmentation may be most beneficial during activities that produce predictable sounds that do not already dominate users’ attention.

In addition to acoustic characteristics of activities, our study identified that an activity’s cognitive demand also affected the experience of using audio augmentation in mindfulness practice. Cognitively demanding tasks could limit attentional capacity for engaging with the amplified sounds, reducing the extent to which such augmentation can support mindful awareness. This finding is consistent with prior literature, which encourages practicing mindfulness during familiar activities, such as cooking, brushing one’s teeth, and folding clothes~\cite{hanh2016miracle}, in which the routine nature affords users with unoccupied cognitive bandwidth. Our participants’ experiences further suggested, however, that cognitive demand depended not only on the task itself but also on the person and context. Some participants chose to use the device during computer work because it fit their routines, particularly when the work involved housekeeping tasks. On the other hand, cooking may demand substantial attention from someone who performs it as a career. These findings broaden our understanding of types of everyday activities that might benefit from audio augmentation as mindfulness support.

\subsection{Roles of Sensory Technology in Everyday Mindfulness Trainings}
Beyond identifying factors that influence the use of audio augmentation in everyday mindfulness, our diary study allowed us to examine the roles the system played in supporting practice. These insights can inform different pathways for sensory technologies to augment existing mindfulness coaching.

Participants’ qualitative accounts suggest that audio augmentation made it easier to enter a mindful state during everyday activities. They suggested that this support could be particularly valuable for beginners, who may struggle to see improvement and stay motivated for continued practice. These findings point to the potential role of sensory technology to support engagement in mindfulness training. Prior work highlights that consistent engagement with practice is an important factor in realizing the long-term benefits of mindfulness~\cite{strohmaier2020relationship}, while sustained engagement remains a challenge in existing mindfulness tools~\cite{linardon2023rates}. Our findings open up pathways for future work to examine whether ease of entering a mindful state enabled by an audio augmentation device leads to more frequent and sustained mindfulness practice in everyday life.

Participants further suggested that audio augmentation could provide an experiential target for independent practice. In this context, the system helped them recognize what sustained awareness of those sounds might feel like. One way to integrate the system into a mindfulness training program is therefore as an experiential demonstration before independent practice. Practitioners could then reflect on where they directed their attention and try attending to similar sounds without the device. Prior work on auditory cueing provides theoretical support for such usage, as drawing attention to particular sound features can make those features easier to notice later in a complex mixture~\cite{hafter1993attending, schlauch1991listening}. More recent work has explored this approach with the textures of sounds produced during everyday activities~\cite{tao2026reperceiving}. Our findings suggest the possibility of audio augmentation to support users in learning what to listen for in mindfulness practice beyond device use. 

Beyond easing transition into mindfulness and providing experiential targets for practice, participants also highlighted the device's potential benefits for experienced practitioners. For those who already have an established mindfulness routine, the device could support the deepening of their practice. Rather than relying on the device for structured guidance, these practitioners could adapt the audio augmentation to contexts they found personally useful. For example, participants described using the device to identify new mindfulness anchors that they might not have considered before, as well as to practice mindfulness during more challenging activities. These accounts suggest that audio augmentation could serve not only as an entry point into mindfulness but also as a tool for extending and deepening existing practices.

Despite these potential roles of sensory technology in supporting mindfulness training, participants cautioned that technological tools cannot replace the intention and deliberate effort required to remain mindful. Prior work identified intention and attention as interwoven components of mindfulness~\cite{shapiro2006mechanisms} and argued that technology should not shift responsibility for practice away from practitioners~\cite{parsons2019designing}. We therefore encourage future work to regard sensory technology as a scaffold: temporary support that eases entry into practice or provides an experiential reference point, while leaving the work of achieving awareness with the practitioners.

\subsection{Limitations}
This study serves as a first step in exploring audio augmentation of manual interactions to support mindfulness in out-of-lab settings and has several limitations.  

First, our sample size is relatively small and does not capture the full range of experiences of possible mindfulness practitioners. Specifically, our participant pool was primarily composed of mindfulness practitioners aged 25 to 35, which may limit the generalizability of the findings to younger and older practitioners. Moreover, the requirement for in-person pickup of the device kit constrained recruitment to participants residing in the local area of the institution. As future work, researchers should extend the evaluation to a more diverse participant sample and consider how the system affects mindfulness practitioners across different demographics and levels of mindfulness practice  experience.

On the system side, while we strive to make the setup wearable, the current form factor still has limitations in terms of usability. For instance, the lanyard structure used to carry the phone keeps it easily accessible; however, it may dangle or interfere in certain situations, such as when users bend over, which can reduce comfort and ease of use. Moreover, while the wired configuration helps minimize latency introduced by audio streaming, the cables may become tethered or interfere during everyday interactions. Therefore, the current hardware setup might limit participants from exploring a wider range of everyday activities, particularly those involving a lot of movement, such as sports or running. As future work, researchers should explore more wireless hardware configurations. This could be achieved by integrating radio-based audio transmission to support both low-latency performance and improved mobility across contexts.

Our current study design also has several limitations. Although the study aims to capture in-the-moment experiences, sustaining mindfulness is ideally supported without interruptions. For instance, a couple of participants reported that the task of taking photos to document their activities as part of the study could interrupt their ongoing activities and mindful engagement. The awareness of the system logging their app usage may also alter their natural behavior. Additionally, due to privacy considerations, participants were instructed to use the system in private environments. These constraints imposed by the study design may have limited the scope of naturally occurring behaviors captured in the study. Future deployments of the system should consider reducing participant burden and exploring less intrusive methods for capturing user experiences.

\subsection{Future Work}
The insights gained from this paper reveal several avenues for future research.

\textbf{Adaptive Augmentation Based on Characteristics of Natural Sounds.} One promising direction for future research is the development of adaptive approaches for augmenting natural sensory experiences. Unlike system-generated guidance cues, natural sensory cues are inherently dynamic and context-dependent, making a fixed augmentation unlikely to be effective in all situations. Our findings suggest that the impact of audio augmentation depends partly on the acoustic properties of the sounds being augmented. For example, amplification was often more beneficial for subtle sounds that might otherwise go unnoticed than for sounds that already strongly attracted attention. Future research could therefore explore adaptive strategies that adjust augmentation in real time based on factors such as input volume, auditory saliency, and surrounding noise levels.

\textbf{Understanding Efficacy of Audio Augmentation in Scaffolding Mindfulness Training.} Building on the insights from the current diary study, another promising direction is to evaluate the efficacy of audio augmentation for supporting mindfulness training. While participants suggested that the system may lower the effort required to enter a mindful state and promote more consistent practice, its effectiveness in achieving these outcomes remains unclear. Future work could investigate this question through a between-subjects study comparing novice practitioners who receive mindfulness training with and without audio augmentation support. Such a study could assess perceived cognitive load during each mindfulness session, the effectiveness of the training in facilitating mindful states, and participants' adherence to the practice over time. 

\textbf{Extending the Design Space of Augmenting Natural Sensory Experience.} In this paper, we used the augmentation of manual interaction sounds as a basis to understand the in-the-wild use of natural sensory experiences to support everyday mindfulness. The design space of augmenting natural sensory experiences extends beyond hand interaction sounds. An extension of this work would be to enhance naturally occurring sounds produced by other parts of the body, as hand interactions represent only a fraction of everyday interactions. These could include footsteps, bodily movements, and internally generated sounds such as breathing or stomach rumbling. The modality of natural sensory experience could also extend beyond auditory cues. Everyday interactions are fundamentally tactile, while many fine-grained textures of interaction are often not perceptually salient. As such, future work could explore augmenting tactile sensations as a complementary modality to auditory feedback, providing richer access to the subtle qualities of everyday actions.

\section{Conclusion}
In this work, we conducted a week-long diary study with 12 experienced mindfulness practitioners to examine situated use of audio-augmented sensory technology in everyday mindfulness. Participants collectively used the system for nearly 90 hours, contributing 84 diary entries and 12 post-study interviews. We found that participants perceived audio augmentation to support their mindfulness practice, and its usage depended on task demands and the auditory characteristics of the ongoing activities. The system was found to support entering and deepening mindfulness practice, and could be beneficial for both novices and experienced practitioners. This work takes a step toward bringing audio augmentation of manual interactions beyond the laboratory and identifies factors that facilitate its use in everyday mindfulness and potential roles in mindfulness training programs.

\bibliographystyle{ACM-Reference-Format}
\bibliography{reference.bib}

\end{document}